%% file: main.tex
\documentclass[sigconf]{acmart}

\usepackage{algorithmic}
\usepackage{array}
\usepackage{graphicx}
\usepackage{textcomp}
\usepackage{hyperref}
\usepackage{multirow}
\usepackage{url}
\usepackage{subfigure}
\usepackage{xspace}
\usepackage{soul}
\usepackage{color}

\newcommand{\tour}{\textrm{TOUR}\xspace}
\newcommand{\idea}{\textrm{IDEA}\xspace}

\AtBeginDocument{%
  \providecommand\BibTeX{{%
    \normalfont B\kern-0.5em{\scshape i\kern-0.25em b}\kern-0.8em\TeX}}}

\setcopyright{iw3c2w3}
\copyrightyear{2021}
\acmYear{2021}
\acmConference[WWW '21 Companion]{Companion Proceedings of the Web Conference 2021}{April 19--23, 2021}{Ljubljana, Slovenia}
\acmBooktitle{Companion Proceedings of the Web Conference 2021 (WWW '21 Companion), April 19--23, 2021, Ljubljana, Slovenia}
\acmPrice{}
\acmDOI{10.1145/3442442.3458612}
\acmISBN{978-1-4503-8313-4/21/04}

\begin{document}

\title{TOUR: Dynamic Topic and Sentiment Analysis of User Reviews for Assisting App Release}



\author{Tianyi Yang}
\affiliation{
  \institution{The Chinese University of Hong Kong}
  \city{Hong Kong}
  \country{China}}
\email{tyyang@cse.cuhk.edu.hk}

\author{Cuiyun Gao}
\affiliation{
  \institution{Harbin Institute of Technology}
  \country{Shenzhen}
  \country{China}}
\authornote{Cuiyun Gao is the corresponding author.}
\email{gaocuiyun@hit.edu.cn}

\author{Jingya Zang}
\affiliation{
  \institution{Harbin Institute of Technology}
  \country{Shenzhen}
  \country{China}}
\email{zjyzangjingya@163.com}

\author{David Lo}
\affiliation{
  \institution{Singapore Management University}
  \country{Singapore}}
\email{davidlo@smu.edu.sg}

\author{Michael R. Lyu}
\affiliation{
  \institution{The Chinese University of Hong Kong}
  \city{Hong Kong}
  \country{China}}
\email{lyu@cse.cuhk.edu.hk}



\input{sections/abstract.tex}

\ccsdesc[500]{Software and its engineering~Software functional properties}
\ccsdesc[300]{Information systems~Information integration}

\keywords{App review, review topic, sentiment analysis}

\maketitle

\input{sections/intro.tex}
\input{sections/approach.tex}
\input{sections/usage.tex}
\input{sections/case.tex}
\input{sections/literature.tex}

\input{sections/conclu.tex}

\begin{acks}
  The work was supported by the Research Grants Council of the Hong Kong Special Administrative Region, China (CUHK 14210920), and the National Natural Science Foundation of China under project No. 62002084.
\end{acks}

\bibliographystyle{ACM-Reference-Format}
\bibliography{main}


\end{document}

%% file: sections/abstract.tex
\begin{abstract}
  App reviews deliver user opinions and emerging issues (e.g., new bugs) about the app releases.
  Due to the dynamic nature of app reviews, topics and sentiment of the reviews would change along with app release versions.
  Although several studies have focused on summarizing user opinions by analyzing user sentiment towards app features, no practical tool is released.
  The large quantity of reviews and noise words also necessitates an automated tool for monitoring user reviews.
  In this paper, we introduce \textbf{\tour} for dynamic \textbf{TO}pic and sentiment analysis of \textbf{U}ser \textbf{R}eviews.
  \tour is able to (i) detect and summarize emerging app issues over app versions, (ii) identify user sentiment towards app features, and (iii) prioritize important user reviews for facilitating developers' examination.
  The core techniques of \tour include the online topic modeling approach and sentiment prediction strategy. \tour provides entries for developers to customize the hyper-parameters and the results are presented in an interactive way.
  We evaluate \tour by conducting a developer survey that involves 15 developers, and all of them confirm the practical usefulness of the recommended feature changes by \tour. 
\end{abstract}

%% file: sections/intro.tex
\section{Motivation}\label{sec:intro}
App user reviews reflect users' instant experience with the app.
Developers are eager to know users' opinions about their apps after release, including which features are favorable or unfavorable by users, presence of bugs, and desired new requirements.
Timely distilling user reviews could facilitate the app release process for developers, e.g., addressing issues or adding new features in the next release.
For example, Facebook Messenger received massive low ratings in August 2014 and suffered a great loss of users, because the version contained serious privacy issues (e.g., access to users' mobile phone photos and contact numbers)~\cite{facebookexample}.
However, the complaints about the privacy issue by a few users had already surfaced right after the release on Apple's App Store.
The serious issue could be effectively mitigated if it were timely identified from user reviews.

The large volume and noisy characteristics of user reviews~\cite{di2016would} increase the burden of manually checking the reviews version by version.
Existing commercial platforms for app market analytics, such as App Annie~\footnote{\url{https://www.appannie.com/}}, simply list all user reviews without providing in-depth analysis of user reviews. 
Prior studies on automatic review analysis either rely on manually annotated data for training~\cite{DBLP:conf/kbse/GuK15}, which is labor-intensive, or directly adopt common sentiment analysis tools for predicting the user opinions~\cite{DBLP:conf/re/GuzmanM14, DBLP:conf/www/LuizVAMSCGR18}. However, according to Novielli et al.'s study~\cite{DBLP:conf/msr/NovielliGL08}, the common sentiment analysis tools are proven not to perform well for software engineering tasks. Moreover, they do not consider the dynamic nature of mobile apps, i.e., different app versions are associated with different reviews.

Based on our survey of developers (see Section~\ref{sec:survey}), the vast majority said getting a detailed analysis of user reviews is important. Developers demand a tool that can automatically detect the issues and user sentiment about app features along with app releases.

\section{Key Innovations}\label{sec:innovation}

In this paper, we demonstrate \tour, a customizable and automatic tool for dynamic \textbf{TO}pic and sentiment analysis of \textbf{U}ser \textbf{R}eviews. \tour helps app developers automatically track the topic changes and user sentiment about app features along with versions. With an app version chosen, \tour provides a ``glimpse'' for each topic of user reviews in both phrase and sentence levels, and highlights the emerging topics, so that developers can focus on the important ones. \tour also presents the results of the sentiment analysis in an interactive mode. The key innovation of \tour is an approach for lightweight sentiment analysis of emerging issues in app user reviews based on customizable opinion words instead of external sentiment tools.

%% file: sections/approach.tex
\section{Workflow of \tour}\label{sec:metho}

\begin{figure}
  \centering
  \includegraphics[width=\columnwidth]{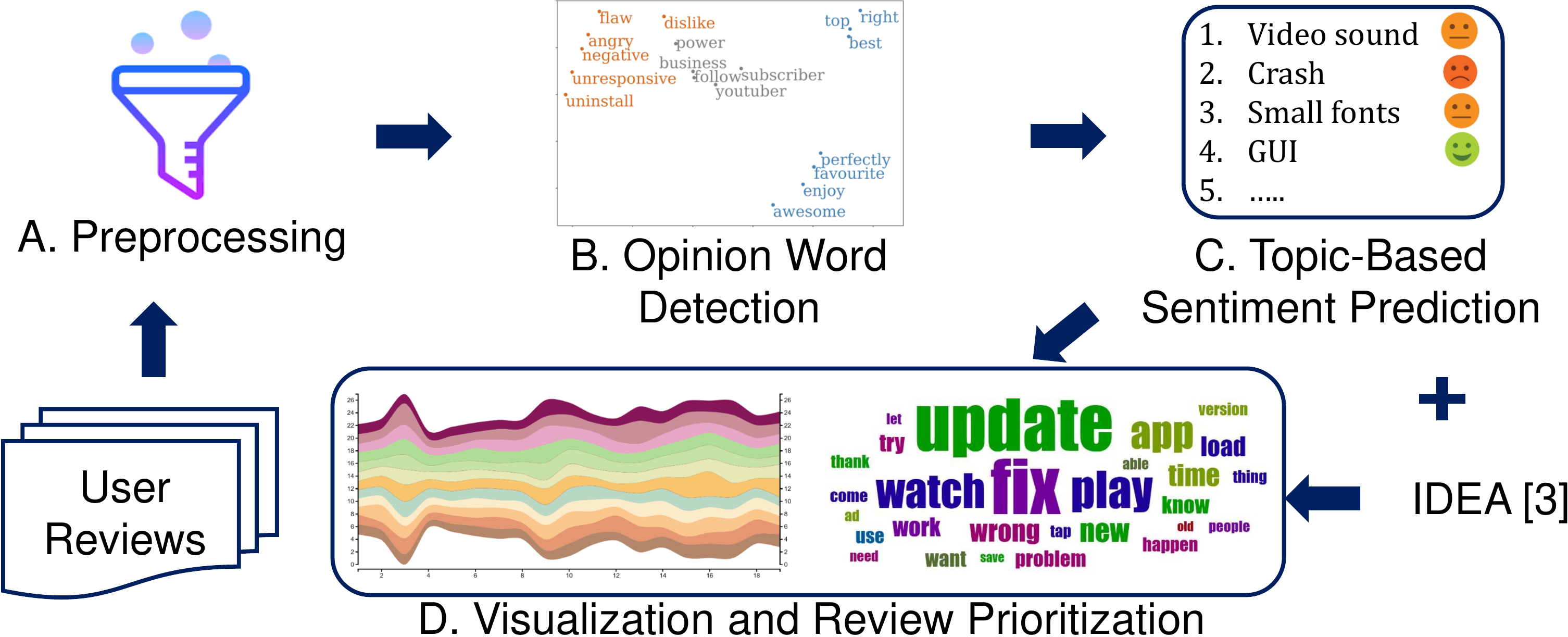}
  \caption{Workflow of \tour.}
  \label{fig:pipeline}
\end{figure}

The workflow of \tour mainly includes four steps, i.e., preprocessing, opinion word detection, topic-based emerging issue detection and sentiment prediction, and visualization and review prioritization, as shown in Figure~\ref{fig:pipeline}.

\subsection{Preprocessing}

The preprocessing step aims at formatting raw data and removing meaningless reviews for subsequent analysis. Following previous work on app review analysis~\cite{DBLP:conf/icse/GaoZD0ZLK19}, we first remove all non-English characters and all non-alphanumeric symbols except the punctuation and then conduct stemming with Porter's Stemmer to convert each word to its root form.
We further clean the review texts based on rules defined in~\cite{DBLP:conf/icse/GaoZD0ZLK19}, such as correcting consecutive duplicates (e.g., \textit{``suuuuper''} is converted into \textit{``super''}), removing consecutively duplicate words (e.g., ``\textit{very very annoying}'' is converted into ``\textit{very annoying}''), and removing all the words whose length is more than 15 (since the word lengths of 99.95\% English words are less than 16\footnote{\url{http://norvig.com/mayzner.html}}). The cleaned reviews are fed into the next step.

\subsection{Opinion Word Detection}

\begin{figure}
  \centering
  \includegraphics[width=0.7\columnwidth]{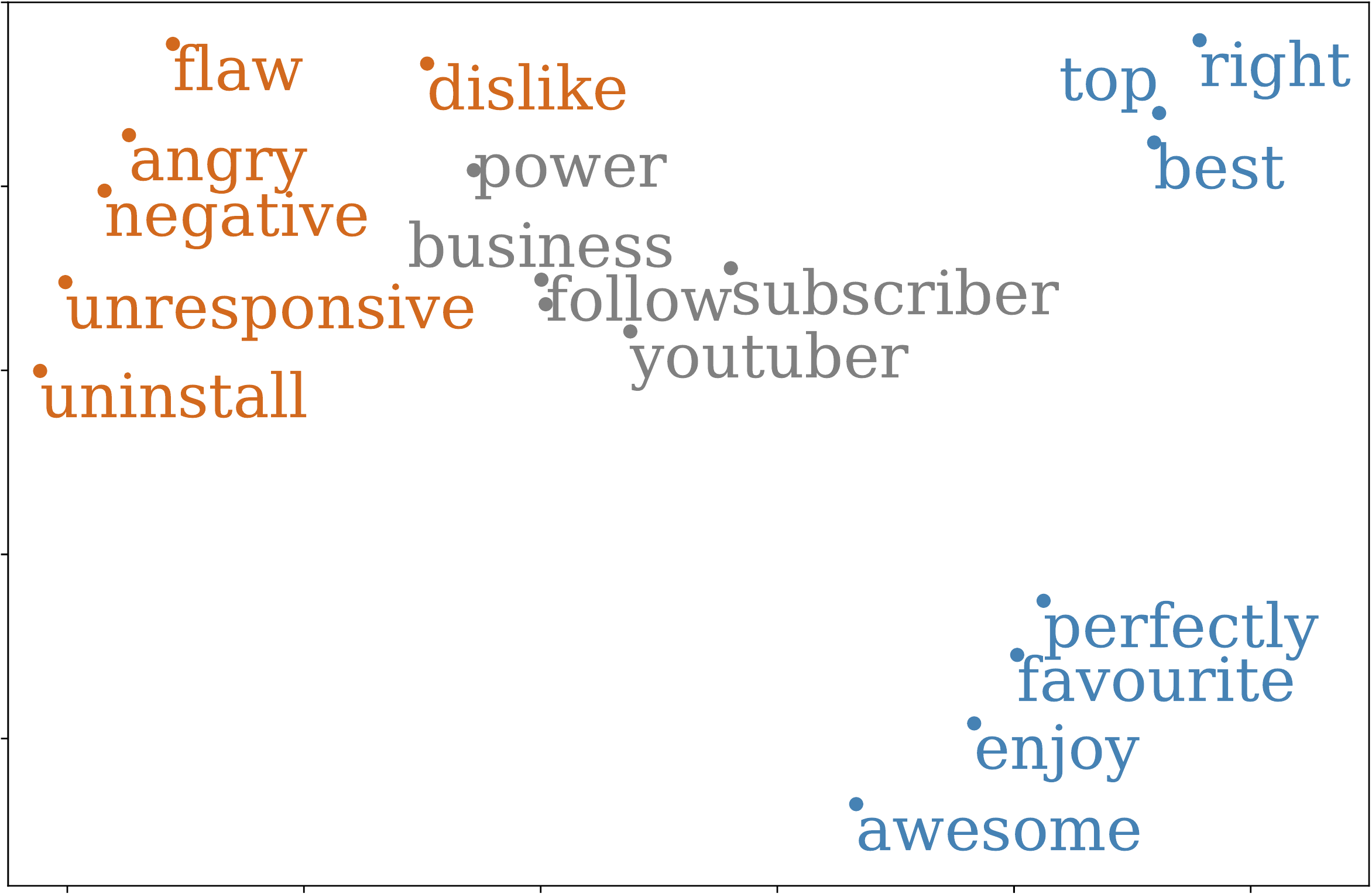}
  \caption{Visualization of word embeddings after dimensionality reduction with t-SNE~\cite{tsne}. The positive opinion words are colored blue and the negative opinion words are colored orange. The aspect words are colored gray.}
  \label{fig:sentiword}
\end{figure}

\begin{figure}[t]
  \centering
  \includegraphics[width=0.95\columnwidth]{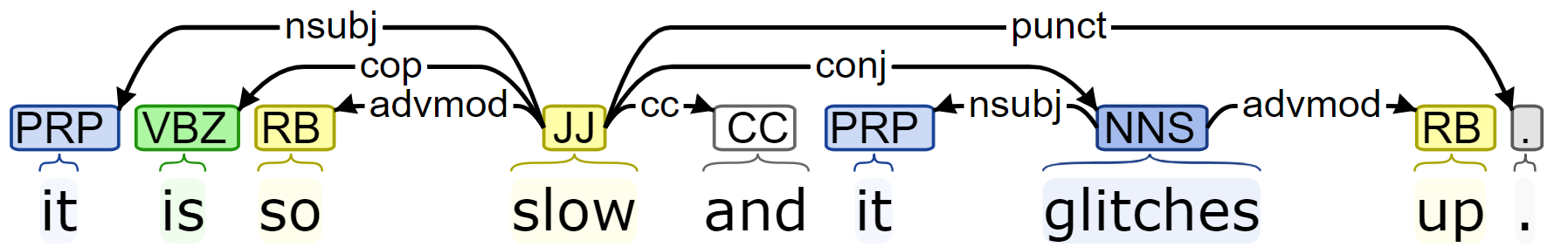}
  \caption{An example of parsed semantic dependency graph.}
  \label{fig:dependency}
\end{figure}

In this step, we aim at identifying opinion words and their sentiment polarities, i.e., negative, neutral, or positive. We identify opinion words by aspect words which usually describe app features and tend to be nouns~\cite{DBLP:conf/kbse/GuK15}. For the review ``\textit{I like Facebook App’s multimedia features but the battery consumption sucks}'', the aspect words are ``\textit{multimedia}'' and ``\textit{battery consumption}''. We then extract opinion words according to the relations with the aspect words, where the relations are captured according to the semantic dependency graph~\cite{cer2010} of the review sentence.
To accurately extract the opinion words, we choose three types of dependency relations, i.e., \emph{noun of subject}, \emph{direct object}, and \emph{adjective modifier}, as listed in Table~\ref{tab:dep_example}. The opinion words are identified if they present one type of the dependency relations with the aspect words.
Figure~\ref{fig:dependency} illustrates an example of the semantic dependency graph for the review text ``\textit{it is so slow and it glitches up}''. We can observe that the opinion words, i.e., ``\textit{slow}'' and ``\textit{glitches}'', present an \texttt{nsubj} (indicating normal subject) relationship with the corresponding aspect word, i.e., ``\textit{it}''.

\begin{table}[t]
  \centering
  \caption{Aspect words and opinion words in different contexts.}
  \label{tab:dep_example}
  \scalebox{0.8}{
    \begin{tabular}{|l|l|l|}
      \hline
      Dependency         & Examples of \textbf{aspect words} and \textit{opinion words} \\
      \hline
      noun of subject    & This \textbf{app} \textit{crashed} on launch.                \\
      direct object      & I \textit{dislike} the \textbf{app}.                         \\
      adjective modifier & Book the \textit{cheapest} \textbf{flight}.                  \\
      \hline
    \end{tabular}
  }
\end{table}

\begin{figure*}[t]
  \centering
  \subfigure[]{
    \includegraphics[width=1.18\columnwidth]{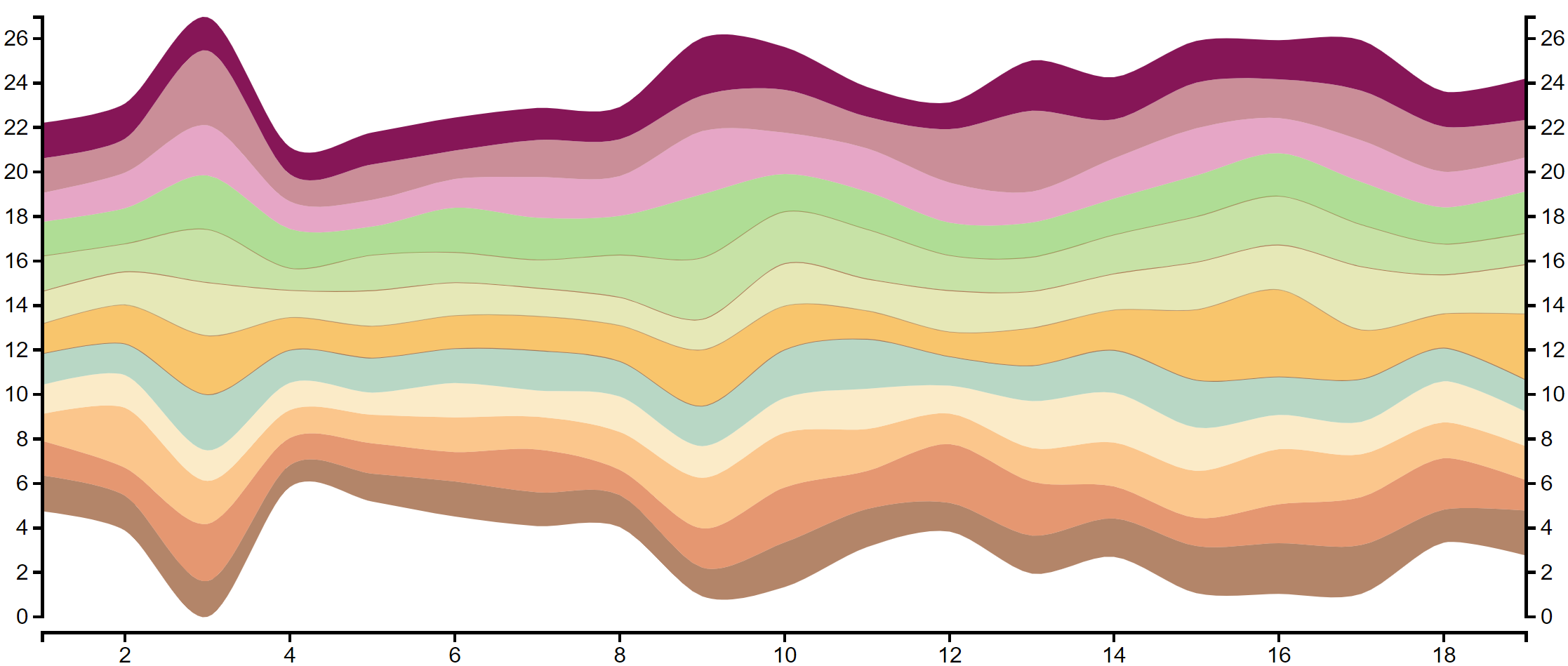}%
    \label{fig:river}
  }
  \hfill
  \subfigure[]{
    \includegraphics[width=0.75\columnwidth]{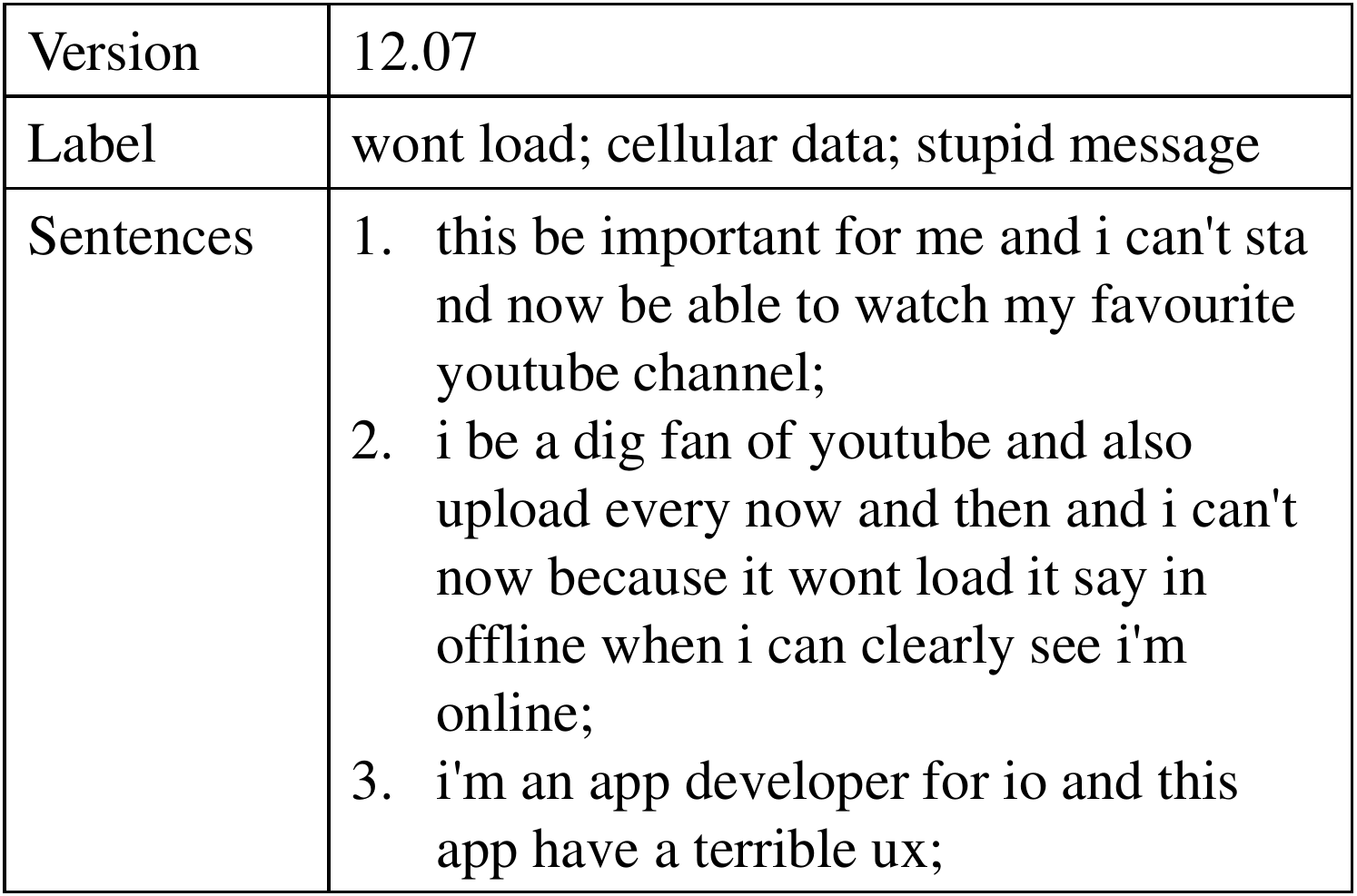}%
    \label{fig:topicdetail}
  }
  \caption{(a) Visualization of issue river. X-axis and Y-axis indicate the version and the width of issue river, respectively. All the topics constitute the issue river, and each branch of the river represents one topic/issue highlighted by a different color. (b) Issue with representative sentences and sentence labels.}
\end{figure*}


To predict the sentiment polarities of the opinion words, prior studies~\cite{DBLP:conf/re/GuzmanM14,DBLP:conf/kbse/GuK15} rely on common sentiment analysis tools that have been shown~\cite{DBLP:conf/msr/NovielliGL08} to be inaccurate for software engineer datasets. Besides, for different apps, the same opinion words can also exhibit totally different sentiment polarities. For example, for the review of YouTube, ``\textit{Can you not make the ad \textbf{louder} than the music videos?}'', the opinion word ``\textit{loud}'' conveys negative sentiment; while for the review of NOAA Radar ``\textit{I really like the feature of sending a \textbf{loud} alert to my phone during pending dangerous weather}'', the word ``\textit{loud}'' is positive in sentiment. To adaptively estimate the sentiment of opinion words, we retrain fine-tuned existing word embeddings~\cite{pennington2014glove} with every app's reviews.
To validate the effectiveness of the retrained embeddings, we conduct dimension reduction on the retrained word embeddings of YouTube with t-SNE~\cite{tsne} and visualize them in Figure~\ref{fig:sentiword}. The positive opinion words are colored blue and the negative opinion words are colored orange. The aspect words are colored gray. Figure~\ref{fig:sentiword} shows that opinion words expressing positive and negative sentiment are separated well.
We then adopt the seed polarity words released by Wilson et al.~\cite{DBLP:journals/coling/WilsonWH09} as the base seed word sets for sentiment prediction. The seed word sets contain manually-labeled context-free polarity words, e.g., ``\textit{great}'' and ``\textit{hate}''. Note that developers can enrich the seed words by manually entering domain-specific polarity words. The sentiment polarities of the opinion words for each app are finally predicted based on the cosine distances with the seed words. Specifically, we define the positive seed word set as $\mathcal{P}$ and the negative one as $\mathcal{N}$. For each opinion word $w$, the sentiment $S(w)$ is defined as:

\begin{equation}\label{equ:1}
  S(w) = \sum_{n\in\mathcal{N}} \frac{Sim(w, n)}{|\mathcal{N}|} - \sum_{p\in\mathcal{P}} \frac{Sim(w, p)}{|\mathcal{P}|},
\end{equation}

\noindent where $S(w)\in[-1, 1]$ and larger sentiment polarities indicate more negative opinions. $Sim(w,p)$ or $Sim(w,n)$ present the cosine distances between the opinion word $w$ and the positive seed word $p\in\mathcal{P}$ or negative seed word $n\in\mathcal{N}$, respectively.

\subsection{Topic-Based Emerging Issue Detection and Sentiment Prediction}

\tour tracks the emerging issues with our previous work, \idea~\cite{DBLP:conf/icse/GaoZLK18}, where emerging issues are detected by version and the topics are automatically labeled.
It utilizes user reviews as input and outputs labeled topics for each version. It employs an online topic modeling approach to generate version-sensitive topic distributions. The emerging topics are then identified based on a typical anomaly detection method. To make the topics comprehensible, \idea labels each topic with the most representative phrases and sentences.
\tour further predicts the sentiment of each labeled topics. Each topic $t\in T$, where $T$ denotes the set of all extracted topics, is represented by a set of topic words $\{w_1^t,w_2^t,\cdots\}$ ranked by the probability distributions. For ensuring the semantic representativeness of the topic words, we consider the top 30 words of each topic during analyzing the topic sentiment. Specifically, \tour first assigns a sentiment label $l$ to each top word $w$, where the sentiment label is defined according to the sentiment score $S(w)$, i.e., $l\in L$, and $L$ = \{Strongly Positive, Positive, Weakly Positive, Slightly Positive, Slightly Negative, Weakly Negative, Negative, Strongly Negative\}. Based on the assigned labels, \tour directly adopts the sentiment label associated with the most top words as the sentiment label of the topic.

\subsection{Visualization and Review Prioritization}

\tour provides an interactive interface for assisting developers monitoring the topic changes and user sentiment towards app features along with app versions. The topic changes are visualized through issue river~\cite{DBLP:conf/icse/GaoZLK18} and user sentiment analysis results are illustrated as word cloud.
\tour also prioritizes the reviews for each topic to facilitate developers' further analysis. The prioritization is based on the probability distributions of the reviews under the topic, with the corresponding topic words highlighted.


%% file: sections/usage.tex
\section{System Demonstration}\label{sec:howto}

\tour is a web application that monitors the topic changes and user sentiment about app features along with app release, where emerging app issues are alerted and reviews of each topic are prioritized for each app release.
An interactive demonstration of \tour will be provided, and the developers can interact with \tour. The following scenarios will be demonstrated.

\subsection{Parameter Selection}

\begin{figure}
  \centering
  \includegraphics[width=0.8\columnwidth]{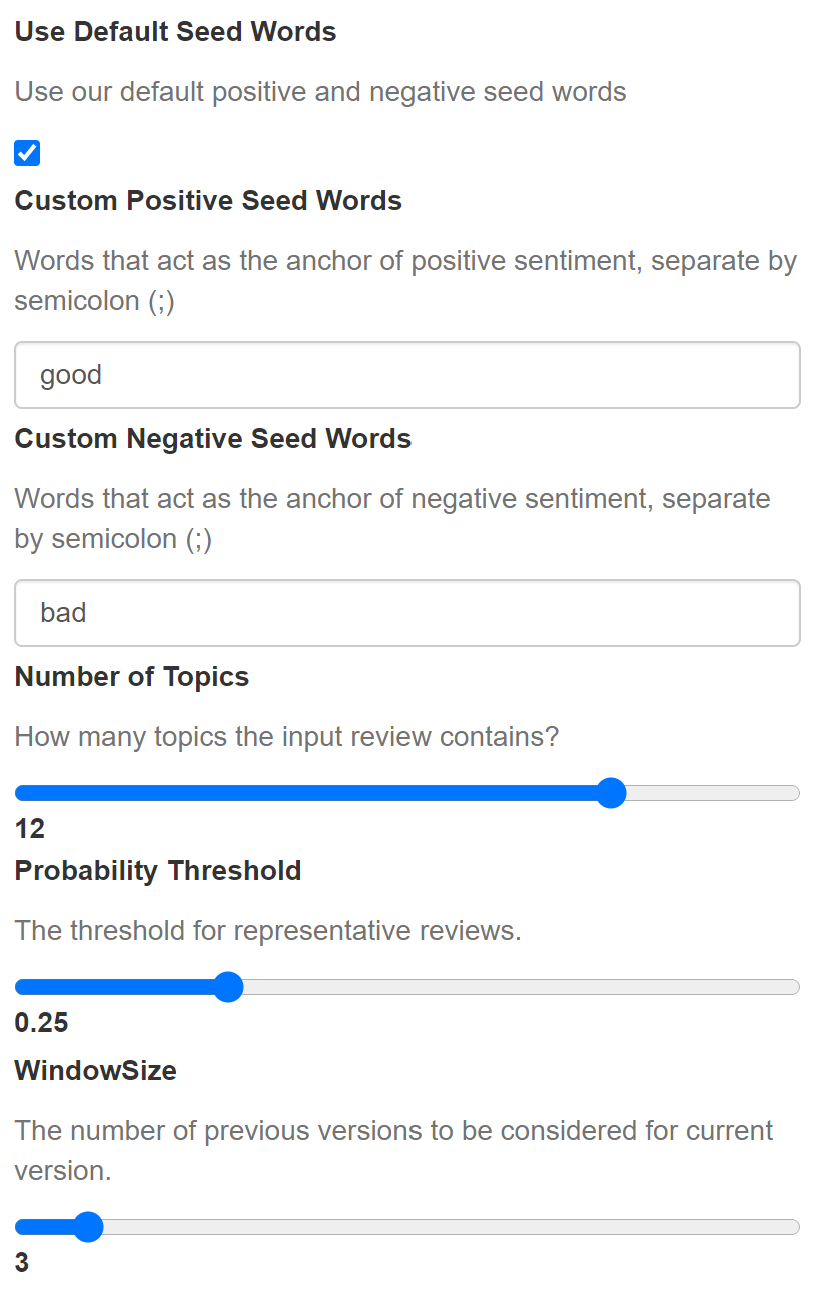}
  \caption{The interface for parameter selection.}
  \label{fig:parameter}
\end{figure}

The input of \tour is a text file where each line is organized as \textit{``[rating]******[review text]******[post date]******[version]******[region]'', using ``******''} to space these review attributes. One example review from YouTube of iOS is \textit{``1.0******Pls fix this. The last update fails to load and play video.******Mar 29, 2017******12.11******SG''}.
As shown in Figure~\ref{fig:parameter}, after uploading reviews, developers will set model parameters including the number of topics, the threshold of the topic probability
for prioritizing reviews of each topic, and other parameters of \idea such as the number of previous versions to be considered for modeling the topics of current version.
\tour provides a comprehensive list of base seed words and developers can enrich seed words with their domain-specific polarity words.

\subsection{Evolution of Topics}

The evolution of topics along with versions will be exposed to the developers through a user interface. The issue river~\cite{DBLP:conf/icse/GaoZLK18} is employed to display topic variations. By moving the mouse on one branch, i.e., one topic, in the issue river, the developers can track the topic changes along with app versions.
Figure~\ref{fig:river} presents an example of the visualized issue river for YouTube iOS.
All the app issues constitute the issue river and each branch of the river indicates one topic, highlighted in a different color.
The topics with wider branches are of greater concern to users.
The width of the $k$-th branch in the $t$-th version is defined as: $ \text{width}_k^t = \sum_a \log Count(a) \times Score_{sen}(a)$, where $Count(a)$ is the count of the phrase label $a$ in the review collection of the $t$-th version, and $Score_{sen}(a)$ denotes the sentiment score of the label $a$.

\subsection{Glimpses of Topics}

\begin{figure}
  \centering
  \includegraphics[width=\columnwidth]{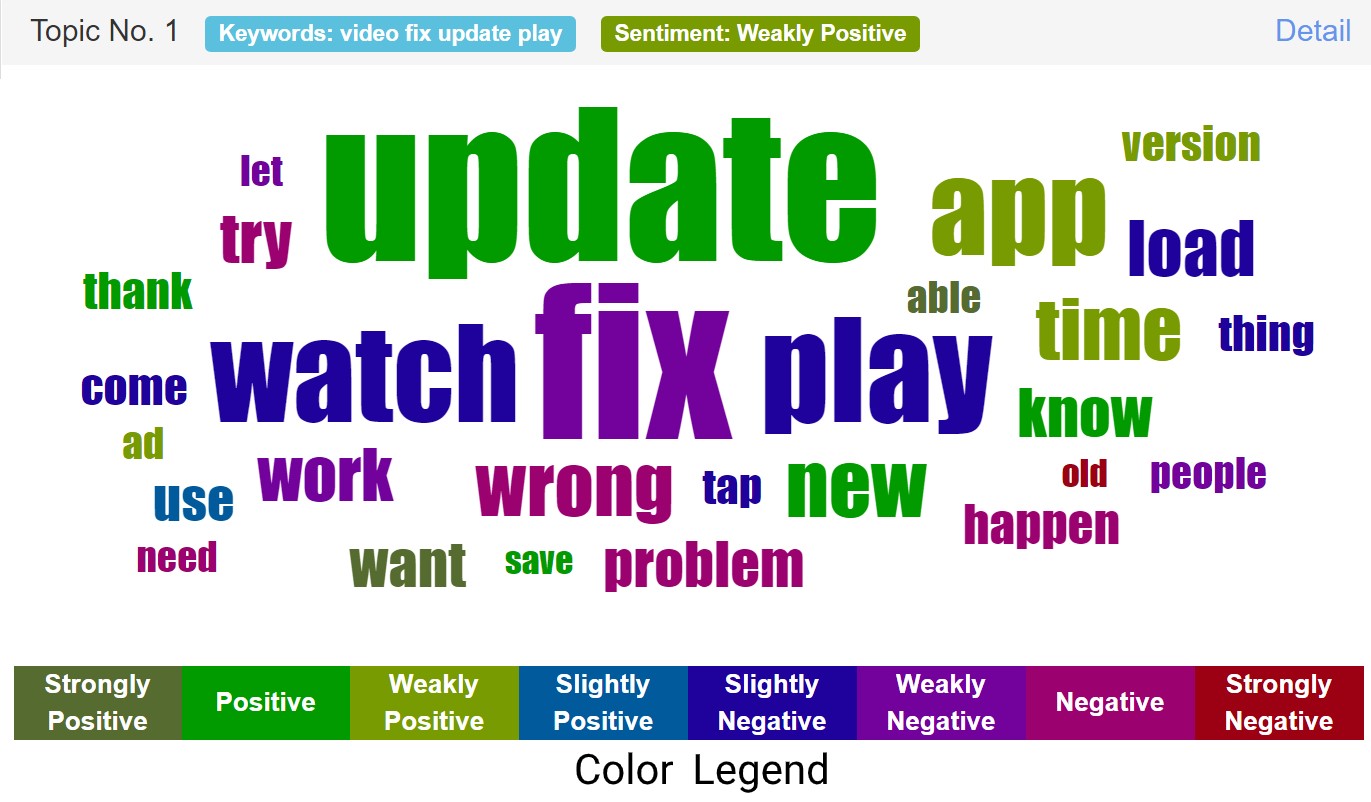}
  \caption{Visualization of issue-sentiment summarization. Font size in the word cloud denotes probability weight of the word in the topic and the color indicates the sentiment label of the word.}
  \label{fig:wordcloud}
\end{figure}

When clicking on one branch, the topic ``glimpse'' will automatically appear, as shown in Figure~\ref{fig:topicdetail}, including representative sentences and sentence labels. For the emerging topics/issues, the sentences are highlighted in yellow for reminding developers. One can also view user sentiment about the topic, visualized in the form of a word cloud, as illustrated in Figure~\ref{fig:wordcloud}. Larger font sizes indicate that the word presents higher probability distribution in the topic and the topic words are displayed in different colors, and the colors indicate the sentiment of the topic words.

\subsection{Prioritized Reviews}

\begin{figure*}
  \centering
  \includegraphics[width=1.8\columnwidth]{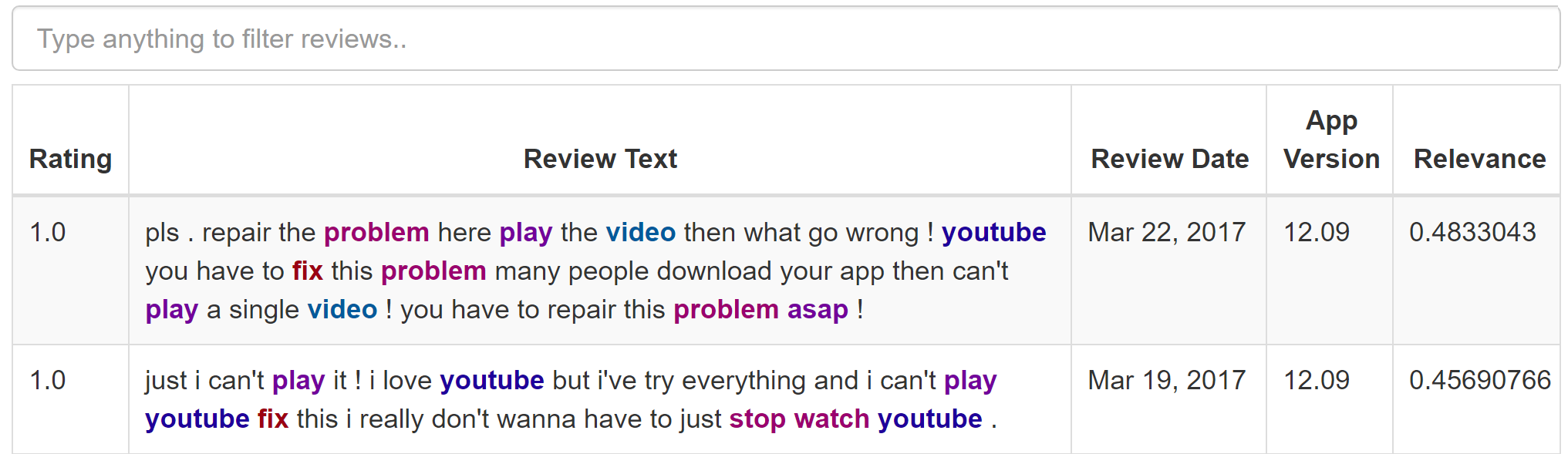}
  \caption{Prioritized associated reviews provided by \tour. The topic words are colorized to indicate the sentiment of the topic words.}
  \label{fig:reviews}
\end{figure*}

Prioritized reviews associated with each topic, as illustrated in Figure~\ref{fig:reviews}. In the ranked review list, review texts and corresponding attributes such as rating, post date, app version, and relevance score (i.e., topic probability distribution) are also displayed for developer's reference. To illustrate users' sentiment in review text, the topic words are highlighted with the same colors as those in the word cloud. The review lists also support full-text retrieval. 
By typing keywords in the search bar, developers can filter reviews by content, rating, and date, and digest user reviews in more details.

%% file: sections/case.tex
\section{Case Study}\label{sec:survey}

In order to evaluate the usefulness of topic-sentiment summary generated by \tour, we conducted an empirical evaluation involving 15 developers.
For the 15 participants, 73\% of them have over one year of software development experience, and 29\% of them have over three years of software development experience.
We adopted the review repository of the six popular apps released by Gao et al.~\cite{DBLP:conf/icse/GaoZLK18}, including YouTube, NOAA Radar, Clean Master, eBay, Swift Keyboard, and Viber. Specifically, we (i) prepared the input reviews for each app; (ii) randomly chose one app and showed the generated issue river, word cloud, and representative reviews to participants; and (iii) invited participants to answer several questions about the usefulness and clarity of the results.

Our GitHub page\footnote{\url{https://yttty.github.io/tour}} depicts the questions and the statistics regarding the answers from the participants.
Overall, 14 out of the 15 participants considered topic summarization and sentiment analysis of app reviews ``important'' or ``very important'', while only one of them insisted that the task was ``not that important''.
All participants agreed that the provided analysis results are useful as a whole, with 46\% of them considered highly useful. Also, the majority of them (13/15) agreed that the visualization are totally comprehensible, while the rest of them partially agreed with that. Most of the participants (14/15) said it is hard to analyze user reviews without \tour.
Additionally, nearly half of them (7/15) declared that our tool saves them at least 50\% of the time, compared to manually analyzing the user reviews.
Additionally, 73\% of them declared that our tool save them over 30\% of the time, compared to manually analyzing user reviews.

Regarding the quality of the results, all of them said that the topic-sentiment summary contains necessary information for review analysis. Besides, 14/15, 15/15, and 15/15 agreed with the usefulness of word cloud visualization, sentiment analysis, and prioritized representative reviews, respectively.
``We pay great attention to user experience and feedback. Your tool is quite helpful.'', a senior developer commented.
To sum up, \tour is helpful for developers and provides useful assistance for user review understanding.

%% file: sections/literature.tex
\section{Related Work}\label{sec:literature}

Existing works~\cite{DBLP:conf/bcshci/IacobVH13, DBLP:conf/kbse/VuNPN15, DBLP:conf/kbse/GuK15, DBLP:conf/www/LuizVAMSCGR18} mainly aim at automatic retrieval of app feature requests from reviews. They mainly analyze static reviews and pay little attention to tracking issue changes.
Gu and Kim~\cite{DBLP:conf/kbse/GuK15} identify reviews related to aspect evaluation based on manually-annotated reviews, and then use the associated reviews for aspect-opinion analysis. 
Luiz et al.~\cite{DBLP:conf/www/LuizVAMSCGR18} also adopts topic modeling to identify app features, but does not consider the emerging issue detection, customizable sentiment seed words, parameter adjustment, and review prioritization involved in our tool.

%% file: sections/conclu.tex
\section{Conclusion}\label{sec:conclu}
App user reviews are valuable for developers but are difficult to analyze.
In this work, we develop an online tool for monitoring topic changes of app reviews along with app release and capturing user sentiment towards app features and issues.
An empirical evaluation shows the effectiveness of \tour for assisting developers in efficiently analyzing user reviews.
In future, we will conduct a more comprehensive empirical evaluation with more developers.

%% file: main.bbl

\begin{thebibliography}{14}


\ifx \showCODEN    \undefined \def \showCODEN     #1{\unskip}     \fi
\ifx \showDOI      \undefined \def \showDOI       #1{#1}\fi
\ifx \showISBNx    \undefined \def \showISBNx     #1{\unskip}     \fi
\ifx \showISBNxiii \undefined \def \showISBNxiii  #1{\unskip}     \fi
\ifx \showISSN     \undefined \def \showISSN      #1{\unskip}     \fi
\ifx \showLCCN     \undefined \def \showLCCN      #1{\unskip}     \fi
\ifx \shownote     \undefined \def \shownote      #1{#1}          \fi
\ifx \showarticletitle \undefined \def \showarticletitle #1{#1}   \fi
\ifx \showURL      \undefined \def \showURL       {\relax}        \fi
\providecommand\bibfield[2]{#2}
\providecommand\bibinfo[2]{#2}
\providecommand\natexlab[1]{#1}
\providecommand\showeprint[2][]{arXiv:#2}

\bibitem[\protect\citeauthoryear{Cer, {de Marneffe}, Jurafsky, and Manning}{Cer
  et~al\mbox{.}}{2010}]%
        {cer2010}
\bibfield{author}{\bibinfo{person}{Daniel Cer},
  \bibinfo{person}{Marie-Catherine {de Marneffe}}, \bibinfo{person}{Daniel
  Jurafsky}, {and} \bibinfo{person}{Christopher~D. Manning}.}
  \bibinfo{year}{2010}\natexlab{}.
\newblock \showarticletitle{Parsing to Stanford Dependencies: Trade-offs
  between speed and accuracy}. In \bibinfo{booktitle}{\emph{7th International
  Conference on Language Resources and Evaluation (LREC 2010)}}.
\newblock
\urldef\tempurl%
\url{http://nlp.stanford.edu/pubs/lrecstanforddeps_final_final.pdf}
\showURL{%
\tempurl}


\bibitem[\protect\citeauthoryear{Di~Sorbo, Panichella, Alexandru, Shimagaki,
  Visaggio, Canfora, and Gall}{Di~Sorbo et~al\mbox{.}}{2016}]%
        {di2016would}
\bibfield{author}{\bibinfo{person}{Andrea Di~Sorbo},
  \bibinfo{person}{Sebastiano Panichella}, \bibinfo{person}{Carol~V Alexandru},
  \bibinfo{person}{Junji Shimagaki}, \bibinfo{person}{Corrado~A Visaggio},
  \bibinfo{person}{Gerardo Canfora}, {and} \bibinfo{person}{Harald~C Gall}.}
  \bibinfo{year}{2016}\natexlab{}.
\newblock \showarticletitle{What would users change in my app? summarizing app
  reviews for recommending software changes}. In
  \bibinfo{booktitle}{\emph{Proceedings of the 24th SIGSOFT International
  Symposium on Foundations of Software Engineering}}. ACM,
  \bibinfo{pages}{499--510}.
\newblock


\bibitem[\protect\citeauthoryear{Gao, Zeng, Lyu, and King}{Gao
  et~al\mbox{.}}{2018}]%
        {DBLP:conf/icse/GaoZLK18}
\bibfield{author}{\bibinfo{person}{Cuiyun Gao}, \bibinfo{person}{Jichuan Zeng},
  \bibinfo{person}{Michael~R. Lyu}, {and} \bibinfo{person}{Irwin King}.}
  \bibinfo{year}{2018}\natexlab{}.
\newblock \showarticletitle{Online app review analysis for identifying emerging
  issues}. In \bibinfo{booktitle}{\emph{Proceedings of the 40th International
  Conference on Software Engineering, {ICSE} 2018}}. \bibinfo{pages}{48--58}.
\newblock


\bibitem[\protect\citeauthoryear{Gao, Zheng, Deng, Lo, Zeng, Lyu, and King}{Gao
  et~al\mbox{.}}{2019}]%
        {DBLP:conf/icse/GaoZD0ZLK19}
\bibfield{author}{\bibinfo{person}{Cuiyun Gao}, \bibinfo{person}{Wujie Zheng},
  \bibinfo{person}{Yuetang Deng}, \bibinfo{person}{David Lo},
  \bibinfo{person}{Jichuan Zeng}, \bibinfo{person}{Michael~R. Lyu}, {and}
  \bibinfo{person}{Irwin King}.} \bibinfo{year}{2019}\natexlab{}.
\newblock \showarticletitle{Emerging app issue identification from user
  feedback: experience on WeChat}. In \bibinfo{booktitle}{\emph{Proceedings of
  the 41st International Conference on Software Engineering: Software
  Engineering in Practice, {ICSE} {(SEIP)} 2019, Montreal, QC, Canada, May
  25-31, 2019}}, \bibfield{editor}{\bibinfo{person}{Helen Sharp} {and}
  \bibinfo{person}{Mike Whalen}} (Eds.). \bibinfo{publisher}{{IEEE} / {ACM}},
  \bibinfo{pages}{279--288}.
\newblock


\bibitem[\protect\citeauthoryear{Gu and Kim}{Gu and Kim}{2015}]%
        {DBLP:conf/kbse/GuK15}
\bibfield{author}{\bibinfo{person}{Xiaodong Gu} {and} \bibinfo{person}{Sunghun
  Kim}.} \bibinfo{year}{2015}\natexlab{}.
\newblock \showarticletitle{What Parts of Your Apps are Loved by Users?}. In
  \bibinfo{booktitle}{\emph{30th {IEEE/ACM} International Conference on
  Automated Software Engineering, {ASE} 2015}}. \bibinfo{pages}{760--770}.
\newblock


\bibitem[\protect\citeauthoryear{Guzman and Maalej}{Guzman and Maalej}{2014}]%
        {DBLP:conf/re/GuzmanM14}
\bibfield{author}{\bibinfo{person}{Emitza Guzman} {and} \bibinfo{person}{Walid
  Maalej}.} \bibinfo{year}{2014}\natexlab{}.
\newblock \showarticletitle{How Do Users Like This Feature? {A} Fine Grained
  Sentiment Analysis of App Reviews}. In \bibinfo{booktitle}{\emph{{IEEE} 22nd
  International Requirements Engineering Conference, {RE} 2014, Karlskrona,
  Sweden, August 25-29, 2014}}. \bibinfo{pages}{153--162}.
\newblock


\bibitem[\protect\citeauthoryear{Iacob, Veerappa, and Harrison}{Iacob
  et~al\mbox{.}}{2013}]%
        {DBLP:conf/bcshci/IacobVH13}
\bibfield{author}{\bibinfo{person}{Claudia Iacob}, \bibinfo{person}{Varsha
  Veerappa}, {and} \bibinfo{person}{Rachel Harrison}.}
  \bibinfo{year}{2013}\natexlab{}.
\newblock \showarticletitle{What are you complaining about?: a study of online
  reviews of mobile applications}. In \bibinfo{booktitle}{\emph{{BCS-HCI} '13
  Proceedings of the 27th International {BCS} Human Computer Interaction
  Conference, Brunel University, London, UK, 9-13 September 2013}}.
  \bibinfo{pages}{29}.
\newblock


\bibitem[\protect\citeauthoryear{Luiz, Viegas, de~Alencar, Mour{\~{a}}o,
  Salles, Carvalho, Gon{\c{c}}alves, and da~Rocha}{Luiz et~al\mbox{.}}{2018}]%
        {DBLP:conf/www/LuizVAMSCGR18}
\bibfield{author}{\bibinfo{person}{Washington Luiz}, \bibinfo{person}{Felipe
  Viegas}, \bibinfo{person}{Rafael~Odon de Alencar}, \bibinfo{person}{Fernando
  Mour{\~{a}}o}, \bibinfo{person}{Thiago Salles},
  \bibinfo{person}{D{\'{a}}rlinton B.~F. Carvalho},
  \bibinfo{person}{Marcos~Andr{\'{e}} Gon{\c{c}}alves}, {and}
  \bibinfo{person}{Leonardo~C. da Rocha}.} \bibinfo{year}{2018}\natexlab{}.
\newblock \showarticletitle{A Feature-Oriented Sentiment Rating for Mobile App
  Reviews}. In \bibinfo{booktitle}{\emph{Proceedings of the 2018 World Wide Web
  Conference on World Wide Web, {WWW} 2018, Lyon, France, April 23-27, 2018}}.
  \bibinfo{publisher}{{ACM}}, \bibinfo{pages}{1909--1918}.
\newblock


\bibitem[\protect\citeauthoryear{Novielli, Girardi, and Lanubile}{Novielli
  et~al\mbox{.}}{2018}]%
        {DBLP:conf/msr/NovielliGL08}
\bibfield{author}{\bibinfo{person}{Nicole Novielli}, \bibinfo{person}{Daniela
  Girardi}, {and} \bibinfo{person}{Filippo Lanubile}.}
  \bibinfo{year}{2018}\natexlab{}.
\newblock \showarticletitle{A benchmark study on sentiment analysis for
  software engineering research}. In \bibinfo{booktitle}{\emph{Proceedings of
  the 15th International Conference on Mining Software Repositories, {MSR}
  2018, Gothenburg, Sweden, May 28-29, 2018}},
  \bibfield{editor}{\bibinfo{person}{Andy Zaidman}, \bibinfo{person}{Yasutaka
  Kamei}, {and} \bibinfo{person}{Emily Hill}} (Eds.).
  \bibinfo{publisher}{{ACM}}, \bibinfo{pages}{364--375}.
\newblock


\bibitem[\protect\citeauthoryear{Pennington, Socher, and Manning}{Pennington
  et~al\mbox{.}}{2014}]%
        {pennington2014glove}
\bibfield{author}{\bibinfo{person}{Jeffrey Pennington},
  \bibinfo{person}{Richard Socher}, {and} \bibinfo{person}{Christopher~D.
  Manning}.} \bibinfo{year}{2014}\natexlab{}.
\newblock \showarticletitle{GloVe: Global Vectors for Word Representation}. In
  \bibinfo{booktitle}{\emph{Empirical Methods in Natural Language Processing
  (EMNLP)}}. \bibinfo{pages}{1532--1543}.
\newblock


\bibitem[\protect\citeauthoryear{Smith}{Smith}{2014}]%
        {facebookexample}
\bibfield{author}{\bibinfo{person}{Dave Smith}.}
  \bibinfo{year}{2014}\natexlab{}.
\newblock \bibinfo{title}{Facebook Messenger is Getting Slammed by Tons of
  Negative Reviews Right Now}.
\newblock
\newblock
\urldef\tempurl%
\url{http://www.businessinsider.com/facebook-messenger-app-store-reviews-are-humiliating-2014-8}
\showURL{%
\tempurl}
\newblock
\shownote{Accessed: 2021-03-24.}


\bibitem[\protect\citeauthoryear{Van~der Maaten and Hinton}{Van~der Maaten and
  Hinton}{2008}]%
        {tsne}
\bibfield{author}{\bibinfo{person}{Laurens Van~der Maaten} {and}
  \bibinfo{person}{Geoffrey Hinton}.} \bibinfo{year}{2008}\natexlab{}.
\newblock \showarticletitle{Visualizing data using t-SNE.}
\newblock \bibinfo{journal}{\emph{Journal of machine learning research}}
  \bibinfo{volume}{9}, \bibinfo{number}{11} (\bibinfo{year}{2008}).
\newblock


\bibitem[\protect\citeauthoryear{Vu, Nguyen, Pham, and Nguyen}{Vu
  et~al\mbox{.}}{2015}]%
        {DBLP:conf/kbse/VuNPN15}
\bibfield{author}{\bibinfo{person}{Phong~Minh Vu}, \bibinfo{person}{Tam~The
  Nguyen}, \bibinfo{person}{Hung~Viet Pham}, {and} \bibinfo{person}{Tung~Thanh
  Nguyen}.} \bibinfo{year}{2015}\natexlab{}.
\newblock \showarticletitle{Mining User Opinions in Mobile App Reviews: {A}
  Keyword-Based Approach {(T)}}. In \bibinfo{booktitle}{\emph{30th {IEEE/ACM}
  International Conference on Automated Software Engineering, {ASE} 2015}}.
  \bibinfo{pages}{749--759}.
\newblock


\bibitem[\protect\citeauthoryear{Wilson, Wiebe, and Hoffmann}{Wilson
  et~al\mbox{.}}{2009}]%
        {DBLP:journals/coling/WilsonWH09}
\bibfield{author}{\bibinfo{person}{Theresa Wilson}, \bibinfo{person}{Janyce
  Wiebe}, {and} \bibinfo{person}{Paul Hoffmann}.}
  \bibinfo{year}{2009}\natexlab{}.
\newblock \showarticletitle{Recognizing Contextual Polarity: An Exploration of
  Features for Phrase-Level Sentiment Analysis}.
\newblock \bibinfo{journal}{\emph{Computational Linguistics}}
  \bibinfo{volume}{35}, \bibinfo{number}{3} (\bibinfo{year}{2009}),
  \bibinfo{pages}{399--433}.
\newblock


\end{thebibliography}
